# A new kind of Hermite-Gaussian-like optical vortex generated by cross-phase

Chen Wang (王 琛)[1], Yuan Ren (任 元)[1,2]*, Tong Liu (刘 通)[1] Linlin Chen (陈琳琳)[1] and Song Qiu (邱松)[1]

[1]*Department of Aerospace Science and Technology, Space Engineering University, Beijing 101416, China*

[2]*State Key Lab of Laser Propulsion & its Application, Space Engineering University, Beijing 101416, China*

*Corresponding author: renyuan_823@aliyun.com;*

We propose a new kind of optical vortex called the Hermite-Gaussian-like optical vortex (HGOV) inspired by the cross-phase (CP). Theoretically, we investigate how the CP is decoupled from the phase of a cylindrical lens. We also investigate the propagation characteristics of an HGOV, which has a Hermitian-Gaussian-like intensity distribution but still retains the orbital angular momentum. Furthermore, we derive the Fresnel diffraction integral of an HGOV and study the purity at infinity. Besides, we show a novel function of the self-measurement of the HGOV. Finally, we show that we can change the relative positions of singularities and the direction of an HGOV precisely, which facilitates applications in optical micro-manipulation.

doi:10.3788/COLXXXXXX.XXXXXX.

Optical vortices (OVs) with the phase factor $\exp(im\phi)$, carry orbital angular momentum (OAM) of $m\hbar$ per photon, where $m$ denotes the topological charges (TCs), $\phi$ denotes the azimuthal angle and $\hbar$ is the Planck constant. Nowadays, OVs are motivating a plethora of applications, such as optical micro-manipulation[1, 2], optical communication[3, 4], high-dimensional quantum entanglement[5], and remote sensing of the angular rotation of structured objects via the optical rotation doppler effect[6, 7].

In 2019, the cross-phase (CP), a new kind of phase structure, has been involved in Laguerre-Gauss (LG) beams that open up a new horizon for generation and measurement of OV[8, 9]. Recently, we investigated a generation and measurement method of high-order OVs via the CP, which has been experimentally achieved[10]. However, we ignored the OAM spectrum, and only judged the generated beams by their intensity and phase distributions in ref. [10]. Lately, we find that the OVs modulated by the CP are not the strictly Hermitian Gaussian (HG) mode due to the abnormal OAM spectrum.

In this letter, inspired by the finding above, we propose a new kind of OV called the *Hermite-Gaussian-like optical vortex* (HGOV) in theory, which has a Hermitian-Gaussian-like intensity distribution but still retains the OAM. HGOVs have a novel function of the self-measurement, which reals the value and sign of TCs, no longer need the interferometry to measure it. Further, HGOVs effectively decouple the phase of a spherical lens from the phase of a cylindrical lens, which allows us to control the HGOV more flexibly, whether at near-field or far-field. HGOVs also have a good depth of field due to relatively stable distributions at far-field, which has a significant meaning for precise 3D optical tweezers. In addition, we can change the relative positions of singularities and the direction of HGOVs precisely, which is of great value in the field of multi-particle manipulation.

The form of the CP $\psi$ in Cartesian coordinates $(x, y)$ is

$$\psi(x, y) = u(x\cos\theta - y\sin\theta)(x\sin\theta + y\cos\theta) \quad (1)$$

where the coefficient $u$ controls the conversion rate, the azimuth factor $\theta$ characterizes the rotation angle of converted beams in one certain plane. It is noteworthy that eq.(1) could be simplified to $\psi'(x,y) = uxy$ when $\theta = 0$. The phase term of a cylindrical lens can be expressed as eq. (2) when settled vertical:

$$\psi_p^{90°} = \exp\left(ix^2 \frac{k}{2f}\right) \quad (2)$$

which can be expanded into:

$$\begin{aligned}\psi_p^{90°} &= \exp\left[\frac{ik}{2f} \times \left(\frac{1}{2}(x^2 + y^2) + \frac{1}{2}(x^2 - y^2)\right)\right] \\ &= \exp\left[\frac{ik}{2f} \times \frac{1}{2}(x^2 + y^2)\right] \\ &\quad \exp\left[\frac{ik}{2f}\left(\frac{\sqrt{2}}{2}x - \frac{\sqrt{2}}{2}y\right)\left(\frac{\sqrt{2}}{2}x + \frac{\sqrt{2}}{2}y\right)\right]\end{aligned} \quad (3)$$

where $k = 2\pi/\lambda$, $f$ denotes the focal length and $\lambda$ denotes the wavelength of the incident light. The term $(x^2 + y^2)/2$ is considered as a spherical lens, and if we take eq.(1) into consideration, we can get:

$$\psi_p^{90°} =$$
$$\exp\left[\frac{ik}{2f} \times \frac{1}{2}(x^2 + y^2)\right] \quad (4)$$
$$\exp\left[\frac{ik}{2f}\left(x\cos\frac{\pi}{4} - y\sin\frac{\pi}{4}\right)\left(x\sin\frac{\pi}{4} + y\cos\frac{\pi}{4}\right)\right]$$

Namely, a vertically placed cylindrical lens can be equivalent to one spherical lens plus one CP, which is rotated by 45 degrees. It is to be noted that the phase of a spherical lens limits the generated beams to shape in the back focal plane (or at far-field), but utilizing the CP to generate HGOVs effectively decouples the phase of a spherical lens from the phase of a cylindrical lens. It allows us to control the HGOV more flexibly, whether at near-field or far-field. Further, HGOVs have a good depth of field due to relatively stable distributions at far-field, which has a significant meaning for precise control of 3D optical tweezers. Besides, due to the split singularities, HGOVs possess a unique advantage that could manipulate multiple particles at the same time. It is noteworthy that we define the light field that meets the far-field condition as the far-field hereinafter.

In fact, the coefficient $u$ in a CP needs to be adjusted to meet the requirements for different initial conditions, which is similar to the cylindrical lens that has the following requirement for the waist radius $\omega$ of the incident light[11]:

$$\omega = \sqrt{\frac{(1+\sqrt{2})f\lambda}{\sqrt{2}\pi}} \quad (5)$$

Thus, if $k/(2f)$ is set to $u$, we can derive the expression of $u$ as:

$$u = \frac{1 + 1/\sqrt{2}}{\omega^2} \quad (6)$$

Without loss of generality, may the form of HGOV with $m=1$:

$$U(x,y,0) = \exp\left(-\frac{x^2+y^2}{\omega^2}\right)\exp(x - iy)\psi'(x,y) \quad (7)$$

According to the Fresnel diffraction integral, when the light field mentioned above propagates a certain distance $z$, the output can be expressed that:

$$E(x,y,z) = \frac{1}{i\lambda z}\exp(ikz)\exp\left(\frac{ik}{2z}(x^2+y^2)\right) \\ \times \mathcal{F}\left[U(x_0,y_0,0)\exp\left(\frac{ik}{2z}(x_0^2+y_0^2)\right)\right] \quad (8)$$

where $(x,y,z)$ denotes the observation plane, $\mathcal{F}$ is the Fourier transform.

Firstly, we would like to introduce the propagation properties of an HGOV from near-field to far-field. Under the condition of the coefficient $\omega = 1\text{mm}$, we simulate the propagation of the HGOV from 0m to 5m, as shown in Fig. 1. As the propagation distance increases, intensity distributions of the HGOV gradually tend to HG distributions. However, phase distributions of the HGOV gradually tend to distributions of an OV. Furthermore, to confirm the specific situation of the phase distributions, we also calculate the OAM spectrums. The OAM spectrum for an arbitrary field $\Psi$ can be calculated with[12, 13]

$$P_m = \frac{C_m}{\sum_{-\infty}^{+\infty} C_n} \quad (9)$$

where

$$C_m = \int_0^\infty \langle a_m(r,\phi,z)^* | a_m(r,\phi,z) \rangle r dr \quad (10)$$

$$a_m(r,z) = 1/(2\pi)^{1/2} \int_0^{2\pi} \Psi(r,\phi,z)\exp(-im\phi)d\phi \quad (11)$$

and $(r,\phi,z)$ denotes the polar coordinate thereof. The calculated OAM spectrums are shown in Fig. 1(a) and the percentage marked in the figures represents the proportion of $m=1$. The increasing mode purity of HGOVs also confirms the fact that the phase gradually changes to the OV of $m=1$, which means the CP only changes the intensity distributions but keeps the most OAM, especially when the HGOV propagates to the far-field. Unfortunately, even if an HGOV propagates to the far-field, the purity still cannot reach 100% according to our simulated results. However, HGOVs have a good depth of field due to relatively stable distributions at far-field, which has a significant meaning for precise 3D optical tweezers.

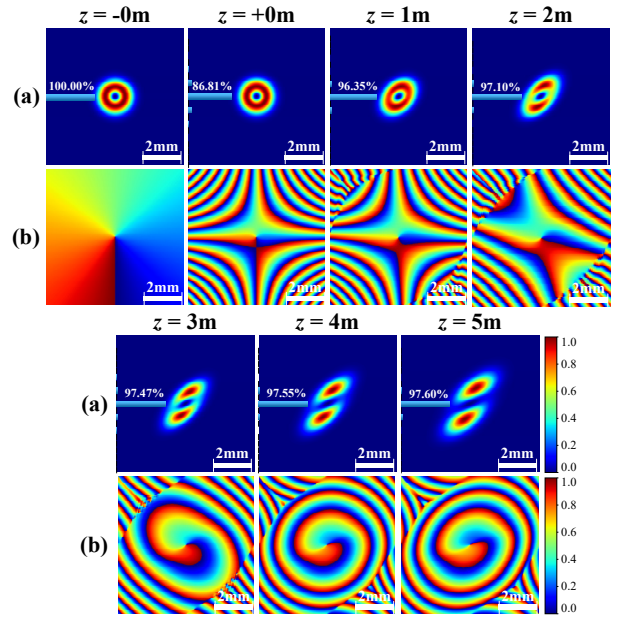

Fig. 1 The simulated propagation of the HGOV from 0m to 5m. z = -0m denotes the original OV with no CP at 0m, and z = +0m denotes the OV immediately with the CP. (a) Simulated intensity

distributions of the HGOV. Blue histograms denote the OAM spectrums, and the percentage marked in the figures represents the proportion of m = 1. (b) Simulated phase distributions are corresponding to (a).

Secondly, we would like to generalize the situation of the HGOV from the far-field to infinity to figure out if there is any possibility that the HGOV can reach the 100% mode purity. Fig. 1 is the numerical simulation based on eq.(7) and eq.(8). In this part, we would like to derive the analytical expressions for the Fresnel diffraction integral after substituting eq.(7) into eq.(8):

$$E(x,y,z) = \frac{1}{i\lambda z}\exp(ikz)\exp\left(\frac{ik}{2z}(x^2+y^2)\right)$$
$$\mathcal{F}\left[\exp(-\frac{x_0^2+y_0^2}{\omega^2})\exp(x_0 - i\,y_0)\right] \quad (12)$$
$$*\mathcal{F}\left[ux_0y_0\exp\left(\frac{ik}{2z}(x_0^2+y_0^2)\right)\right]$$

Using eq.(13) we can simplify eq.(12) to eq.(14):

$$\int_{-\infty}^{\infty} x^n \exp(-px^2+2qx)\mathrm{d}x$$
$$= n!\exp\left(\frac{q^2}{p}\right)\sqrt{\frac{\pi}{p}}\left(\frac{q}{p}\right)^n \sum_{k=0}^{\left[\frac{n}{2}\right]} \frac{1}{(n-2k)!k!}\left(\frac{p}{4q^2}\right)^k \quad (13)$$
$$= \sqrt{\frac{\pi}{p}}\exp\left(\frac{q^2}{p}\right)\sum_{k=0}^{\left[\frac{n}{2}\right]} \frac{n!}{(n-2k)!k!} p^{k-n} 4^{-k} q^{n-2k}$$

$$E(x,y,z) = \frac{\pi}{2\lambda z^2 (\beta\alpha)^{3/2}}(-kx\alpha + U\gamma_1 z)$$
$$\exp\left(ikz + ik\frac{x^2+y^2}{2z} + \frac{\gamma_1^2}{\alpha} - \frac{k^2 x^2}{4\beta z^2}\right)$$
$$+ i\frac{\pi}{2\lambda z^2 (\beta\alpha)^{3/2}}(ky\alpha - U\gamma_2 z) \quad (14)$$
$$\exp\left(ikz + ik\frac{x^2+y^2}{2z} + \frac{\gamma_2^2}{\alpha} - \frac{k^2 y^2}{4\beta z^2}\right)$$

where

$$\begin{cases} \alpha = \left(\frac{1}{\omega^2} + k\frac{1}{2iz} + \frac{U^2}{4\beta}\right) \\ \beta = \left(\frac{1}{\omega^2} + k\frac{1}{2iz}\right) \\ \gamma_1 = \frac{1}{2}\left(k\frac{y}{iz} + k\frac{xU}{2\beta z}\right) \\ \gamma_2 = \frac{1}{2}\left(k\frac{x}{iz} + k\frac{yU}{2\beta z}\right) \end{cases} \quad (15)$$

When $z \to +\infty$, $E(x,y,+\infty) \propto (-x+iy)$, which means that we only get a pure HGOV at infinity. Interestingly, due to decoupling the phase of a spherical lens from the phase of a cylindrical lens, we maintain the consistency of OAM growth rather than decreasing even flipping the sign of OAM[14, 15]. In ref.[15], based on ABCD matrix theory, the total OAM at the initial plane is:

$$\Lambda(0) = \frac{\Phi}{c^2 k} m \quad (16)$$

where $\Phi$ is the total energy flux carried by the beam, $c$ is the light velocity. The OAM after the cylindrical lens is:

$$\Lambda(z) = \frac{\Phi m}{c^2 k} \frac{\left(\frac{z}{z_R}\right)^2 + \left(1-\frac{z}{f_x}\right)^2}{\left(\frac{z}{z_R}\right)^2 + \frac{1}{2}\left(1-\frac{z}{f_x}\right)^2 + \frac{1}{2}} \quad (17)$$

where $z_R = 4k\omega^2$, and $f_x$ denotes the focal length of the cylindrical lens. Although the CP is part of the phase of a cylindrical lens, we can think that the phase of the spherical lens in eq.(3) is close to a plane at infinity. Namely, within the condition that the propagation distance is infinity, we can calculate the OAM of the HGOV by eq.(17): when $z \to +\infty$, $\Lambda(z) \to \Phi m/c^2 k$, which is exactly equal to the OAM at the initial plane. This conclusion confirms the OAM distribution of an HGOV at infinity from the perspective of energy flow.

Nevertheless, HGOVs can still maintain a high mode purity. Furthermore, HGOVs keep the OAM of an OV while retaining the function of the self-measurement, as shown in

Fig. 2. We can directly get the TCs of the HGOVs, which are 2, 3, 4, 5, respectively, no longer need the interferometry to measure it. Meanwhile, the mode purity of simulated HGOVs remains above 97%. Interpretation by phase distributions of

Fig. 2(b), singularities of HGOVs have been split. However, we only concern the total OAM distributions of an HGOV. Besides, due to the split singularities, the HGOV possesses a unique advantage that could manipulate multiple particles at the same time, and we discuss this in detail below.

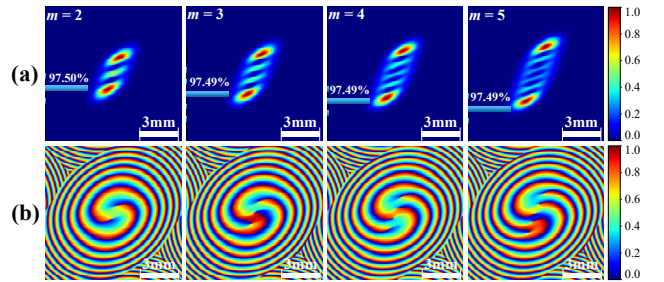

Fig. 2 The function of the self-measurement of the HGOV at far-filed. (a) Simulated intensity distributions of the HGOV of $m$ = 2, 3, 4, 5, respectively. Blue histograms denote the OAM spectrums, and the percentage marked in the figures represents the proportion of $m$ = 2, 3, 4, 5, respectively. (b) Simulated phase distributions are corresponding to (a).

We also can use the function of the self-measurement to achieve the self-sign-measurement of the HGOV, as shown in Fig. 3. Generated HGOVs have different distributions when the TCs have the same value but an opposite sign at far-field.

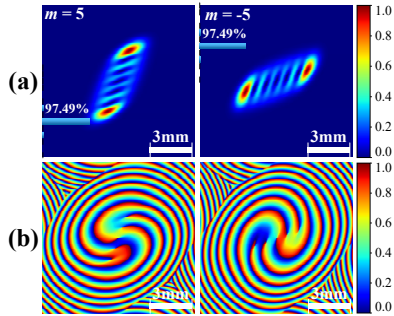

Fig. 3 The self-sign-measurement of the HGOV. (a) Simulated intensity distributions of the HGOV of $m$ = 5, -5, respectively. (b) Simulated phase distributions are corresponding to (a).

We would like to discuss the influence of the coefficient $u$ on the generation of HGOVs. The coefficient $u$ controls the conversion rate of an HGOV to the intensity distribution of the HG beam. However, we find that the mode purity decreases from 99.25% to 99.71 as the coefficient $u$ increases from $0.5u$ to $2.0u$ at far-field, as shown in Fig. 4. Further, we mark the locations of the singularities with the white crosses. It is to be noted that we can change the relative positions of singularities precisely by altering the parameter $u$, which is of great value in the field of multi-particle manipulation.

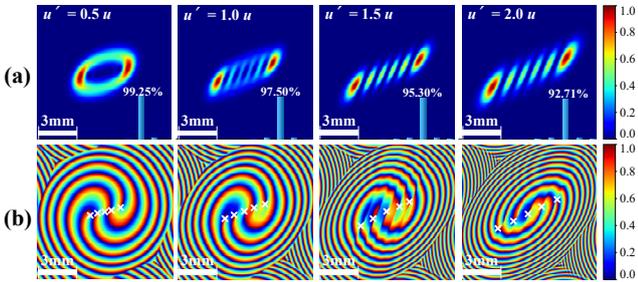

Fig. 4 Simulated distributions of HGOVs with different $u$: $u'$ = $0.5u$, $1.0u$, $1.5u$, $2.0u$, respectively. (a) Simulated intensity distributions. Blue histograms denote the OAM spectrums, and the percentage marked in the figures represents the proportion of $m$ = -5. (b) Simulated phase distributions are corresponding to (a). The white crosses mark the locations of the singularities.

However, it is not enough to change the relative positions of singularities. We also need to accurately control the direction of an HGOV if we want to achieve the multi-particle manipulation at any position. We know that the azimuth factor $\theta$ characterizes the rotation angle of an HGOV in one certain plane in eq.(1), makes it possible to achieve the multi-particle manipulation anywhere. By altering the azimuth factor $\theta$, we simulate HGOVs at $0°$, $45°$, $90°$ and $135°$, respectively, as shown in Fig. 5.

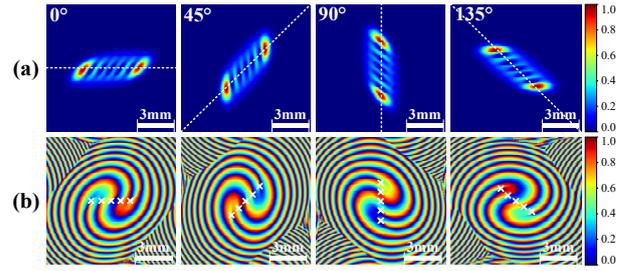

Fig. 5 Simulated results of HGOVs of $m$ = 5 at $0°$, $45°$, $90°$ and $135°$, respectively. (a) Simulated intensity distributions. The white dotted lines are the axis of symmetry of HGOVs. (b) Simulated phase distributions are corresponding to (a). The white crosses mark the locations of the singularities.

In summary, we propose a new kind of OV called the *Hermite-Gaussian-like optical vortex*. Firstly, we show how the CP is decoupled from the phase of a cylindrical lens, which allows us to control the HGOV more flexibly, whether at near-field or far-field. Secondly, we investigate the propagation characteristics of HGOVs, which has a Hermitian-Gaussian-like intensity distribution but still retains the OAM at far-field. HGOVs have a good depth of field due to relatively stable distributions at far-field, which has a significant meaning for precise control of 3D optical tweezers. Theoretically, we derived the diffraction integral formula of HGOVs, which confirms that pure HGOVs only exist at infinity. We also confirm this conclusion from the perspective of energy flow. Thirdly, we introduce a novel function of HGOVs that the function of the self-measurement, which reals the value and sign of TCs, no longer need the interferometry to measure it. In the discussion part, we discuss the influence of the coefficient $u$ on the generation of HGOVs and find that the mode purity decreases as the coefficient $u$ increases at far-field. In addition, we show that we can change the relative positions of singularities and the direction of HGOVs precisely, which is of great value in the field of multi-particle manipulation.

This work was supported in part by the National Nature Science Foundation of China under Grant 11772001 and 61805283, and in part by the Beijing Youth Top-Notch Talent Support Program under Grant 2017000026833ZK23.